\documentclass[12pt]{article}
\usepackage{amsmath,amsfonts}
\usepackage{mathrsfs}
\usepackage{graphicx}

\title{Parameterized mixed cluster editing via modular decomposition\footnote{This work has been partially supported by the
 Conselho Nacional de Desenvolvimento Cient\'{\i}fico e Tecnol\'{o}gico (CNPq), Funda\c c\~ao de Amparo \`a Pesquisa do
 Estado do Rio de Janeiro (FAPERJ) and Coordena\c c\~ao de Aperfei\c coamento do Pessoal de Ensino Superior (CAPES),
 Brazilian research agencies.}}

\author{Maise Dantas da Silva\footnote{P\'{o}lo Universit\'{a}rio de Rio das Ostras, Universidade Federal Fluminense, RJ, Brazil.
E-mail: {\tt maise@vm.uff.br}} \and F\'abio Protti\footnote{Instituto de Computa\c c\~ao, Universidade Federal Fluminense, Rua Passo da P\'{a}tria 156, 24210-240, Niter\'oi,
RJ, Brazil. E-mail: {\tt fabio@ic.uff.br}} \and  Jayme Luiz
Szwarcfiter\footnote{Instituto de Matem\'atica, N\'ucleo de
Computa\c c\~ao Eletr\^onica, and COPPE-Sistemas, Universidade
Federal do Rio de Janeiro, Caixa Postal 68511, 21945-970, Rio de
Janeiro, RJ, Brazil. E-mail: {\tt jayme@nce.ufrj.br}}}

\def\qed{ \fbox \par \medskip }

\def\qed2{ \hfill \framebox[.09in] \par \bigskip }

\def\M{{\cal M}}
\def\L{{\cal L}}
\def\KL{{\cal K}_{\ell}}

\begin{document}

\newtheorem{teo}{Theorem}
\newtheorem{lema}[teo]{Lemma}
\newtheorem{cor}[teo]{Corollary}
\newtheorem{prop}[teo]{Proposition}
\newtheorem{obs}[teo]{Remark}
\newtheorem{defin}[teo]{Definition}

\maketitle

\begin{abstract}

In this paper we introduce a natural generalization of the well-known problems {\sc Cluster Editing} and {\sc Bicluster Editing}, whose parameterized versions have been intensively investigated in the recent literature. The generalized problem, called {\sc Mixed Cluster Editing} or $\M${\sc -Cluster Editing}, is formulated as follows. Let $\M$ be a family of graphs. Given a graph $G$ and a nonnegative integer $k$, transform $G$, through a sequence of at most $k$ edge editions, into a target graph $G'$ with the following property: $G'$ is a vertex-disjoint union of graphs $G_1, G_2, \ldots$ such that every $G_i$ is a member of $\M$. The graph $G'$ is called a {\em mixed cluster graph} or $\M${\em -cluster graph}. Let ${\cal K}$ denote the family of complete graphs, $\KL$ the family of complete $\ell$-partite graphs ($\ell \geq 2$), and
$\L={\cal K} \cup \KL$. In this work we focus on the case $\M = \L$. Using modular decomposition techniques previously applied to {\sc Cluster/Bicluster Editing}, we present a linear-time algorithm to construct a problem kernel for the parameterized version of $\L${\sc -Cluster Editing}.
\end{abstract}

\section{Introduction} \label{sec1}

Edge edition (or edge {\em modification}) problems have been intensively studied within
the context of parameterized complexity theory. The general
formulation for this class of problems is: ``transform an input
graph $G$ into a member of a target family
by editing at most $k$ of its edges.'' For a detailed study on
edge edition problems, see \cite{NSS99}.

In particular, cluster editing problems appeared as a promising
field for this research, due to their applications in
computational biology, data mining, facility location, network
models, etc. For this class of problems, the target family
is usually formed by graphs consisting of a vertex-disjoint union
of cliques ({\sc Cluster Editing}), bicliques ({\sc Bicluster Editing}), or other types of dense and/or regularly structured
graphs. Several recent works have presented results on cluster editing problems,
see for instance~\cite{TBGEP2004,CM2011,DSPS2006,DSPS2009,FLRS2007,GGHN2003,G2007,YZ2007,SST2004}.

A natural generalization of cluster editing problems consists of
defining the target family to contain {\em mixed cluster graphs}. A mixed cluster graph is
a vertex-disjoint union of graphs $G_1, G_2, \ldots$
such that each $G_i$ is a member of a fixed family $\M$.
In this formulation, {\sc Cluster Editing} corresponds precisely to
$\M = {\cal K} = \{K_n \mid n > 0\}$, and {\sc Bicluster Editing} to
$\M = \{K_1\} \cup \{K_{m,n} \mid mn > 0\}$. Let us call such
a generalized problem {\sc Mixed Cluster Editing} or $\M${\sc -Cluster Editing}.
Mixed cluster graphs are also called $\M${\em -cluster graphs}.

The proposed generalization covers the case in which $\M$
includes graphs of two or more well-known families. For a fixed integer
$\ell \geq 2$, define $\KL$ as the family of {\em
$\ell$-cliques}\footnote{In the literature, `$\ell$-clique' also stands for a clique of size $\ell$,
but we employ here the above terminology in order to generalize
the term `bicliques' (for which $\ell=2$).}, consisting of the connected\footnote{A non-trivial edgeless graph is complete $\ell$-partite (with one non-empty color class
and $\ell-1$ empty color classes), but is not connected; thus, according to our definition, it is not an $\ell$-clique.}, complete $\ell$-partite
graphs.
Clearly, $\KL \subseteq {\cal K}_{\ell+1}$, for every $\ell \geq 2$.
Let $\L={\cal K} \cup \KL$.
In this work, we focus on the case $\M=\L$,
that is, the target graph must be a vertex-disjoint union of graphs $G_1, G_2, \ldots$
such that each $G_i$ is a clique or an ${\ell}$-clique.

Since the family $\L$ can be characterized by a finite set of forbidden induced subgraphs
with at most $\ell+2$ vertices
(Proposition~\ref{forbidden}), the tractability of the parameterized version of
$\L${\sc -Cluster Editing}, denoted by $\L${\sc-Cluster Editing}$(k)$,
follows directly from a result by Cai~\cite{C1996}, which provides an $O({(\ell+2)}^{2k}n^{\ell+3})$-time
algorithm to solve it. In fact, Cai's result can also be applied
to $\M${\sc -Cluster Editing} whenever $\M$ is characterized
by a finite set of forbidden induced subgraphs.

We propose a linear-time kernelization algorithm for $\L${\sc -Cluster Editing}$(k)$
that builds a problem kernel with $O(\ell k^2)$ vertices.
Considering the trivial $O({((\ell+2)(\ell+1)/2)}^k)$ time
bounded search tree \cite{NR2000}, this gives an
$O({((\ell +2)(\ell +1)/2)}^k + n + m)$ time algorithm for $\L${\sc -Cluster Editing}$(k)$.

Our kernelization algorithm is based on the modular decomposition techniques
previously applied to {\sc Cluster/Bicluster
Editing}~\cite{DSPS2006,DSPS2009}, extending their usefulness to solve
cluster editing problems in general.
Recent algorithms~\cite{CM2011,G2007,YZ2007} construct kernels for {\sc Cluster Editing} with size $O(k)$, but not in linear time.

The remainder of this work is organized as follows. Section
\ref{sec2} contains basic definitions, notation and preliminary results.
In Section \ref{sec3} we deal with the concept of {\em quotient graphs} and show how it allows us to derive
useful bounds and reduction rules for the kernelization algorithm.
In Section \ref{sec4} we show how to construct a problem
kernel in linear time for $\L${\sc -Cluster Editing}$(k)$.
Finally, Section \ref{sec5} discusses how the kernelization algorithms developed here and in \cite{DSPS2006,DSPS2009}
can be applied to obtain reduced graphs with $O(k)$ vertices, in linear time, both for {\sc Cluster Editing($k$)} and {\sc Bicluster Editing($k$)}.

\section{Preliminaries} \label{sec2}

Let $G$ denote a finite graph, without loops and multiple edges.
If $H$ is an induced subgraph of $G$ then we say that $G$ {\it contains} $H$, or $H$ {\it is contained in} $G$.
The vertex set and the edge set of $G$ are denoted by $V(G)$ and $E(G)$, respectively.
Assume $|V(G)|=n$ and $|E(G)|=m$. A chordless path with $n$ vertices is denoted by $P_n$.
A \emph{clique} is a complete (sub)graph.  A \emph{cluster graph} is a vertex-disjoint union of cliques.
A clique with $n$ vertices is denoted by $K_n$. ${\cal K}$ denotes the family of complete graphs.
A graph is $\ell$-partite if it is $\ell$-colorable.
An \emph{$\ell$-clique} is a connected, complete $\ell$-partite (sub)graph.
$\KL$ denotes the family of $\ell$-cliques, and $\L$ is defined as $\L={\cal K} \cup \KL$.
A $\KL$\emph{-cluster graph} is a vertex-disjoint union of $\ell$-cliques.
An $\L$\emph{-cluster graph} is a vertex-disjoint union of cliques and/or $\ell$-cliques.

We remark that a graph $G$ is a cluster graph if and only if $G$ does not contain $P_3$,
and an $\ell$-cluster graph if and only if $G$ does not contain any of the graphs
$P_4$, $\overline{P_3 \cup K_1}$ and $K_{\ell+1}$ (the graph $\overline{P_3 \cup K_1}$
is called {\em paw}). Denote by $K_r-e$ the complete graph with $r$ vertices minus one edge.
The following proposition characterizes $\L$-cluster graphs by means of forbidden induced subgraphs:

\begin{prop} \label{forbidden}
A graph $G$ is an $\L$-cluster graph if and only if $G$ does
not contain any of the graphs $P_4$, $\overline{P_3 \cup K_1}$ and $K_{\ell+2}-e$.
\end{prop}


\noindent
\emph{\textbf{Proof:}} If $G$ is an $\L$-cluster graph then it is clear that $G$ cannot contain
any of the graphs $P_4$, $\overline{P_3 \cup K_1}$ and $K_{\ell+2}-e$. Conversely, assume that
$G$ does not contain such graphs. Since $G$ contains no $P_4$,
$G$ is a cograph~\cite{CLB81}. Let $H$ be a connected component of $G$. By properties of modular decomposition, $H$ is formed
by disjoint subgraphs $H_1, H_2, \ldots, H_q$ such that each $H_i$ is either trivial or disconnected,
and every vertex of $H_i$ is adjacent to every vertex of $H_j$ for $i \neq j$.
If every $H_i$ is trivial, $H$ is a clique. Otherwise, assume $|V(H_1)| \geq 2$. If $H_1$ contains an
edge $ab$, we can choose a vertex $c$ in a connected component of $H_1$ not containing $ab$ and a vertex $d \in V(H_2)$
to form an induced paw, a contradiction. This means that every $H_i$ is an edgeless graph. To conclude the proof,
since $G$ contains no $K_{\ell+2}-e$,  we have $q \leq \ell$, that is, $H$ is an $\ell$-clique. Hence,
$G$ is an $\L$-cluster graph.  \qed2

An \emph{edition set} $F$ is a set of unordered pairs of
vertices, each pair marked $-$ or $+$, such that:

\begin{itemize}

\item $-ab$ represents the deletion from $E(G)$ of the edge $ab$ \ ({\em edge deletion});

\item $+ab$ represents the addition to $E(G)$ of the edge $ab$ \ ({\em edge addition}).

\end{itemize}

In both cases, we say that $-ab$ or $+ab$ is an {\em edge edition}.
Assume that $F$ does not contain a pair $-ab$ (resp. $+ab$) if $ab \notin E(G)$ (resp. $ab \in E(G)$).
Assume also that no edge is edited more than once in $F$.

Sometimes, the type of edition ($-$ or $+$) will be omitted for simplicity; in this case,
we will denote an edge edition involving vertices $a$ and $b$ simply by $ab$.

We say that an induced subgraph $H$ of $G$ is {\em destroyed} by the edition
set $F$ if there exist $a,b \in V(H)$ such that:

\begin{itemize}

\item  if $ab \in E(H)$ then $F$ contains $-ab$;

\item  if $ab \notin E(H)$ then $F$ contains $+ab$.

\end{itemize}

In either case, we also say that $H$ is destroyed by the corresponding edge edition ($-ab$ or $+ab$).

In this work we are mainly concerned with the following objective: given a graph $G$,
find an edition set $F$ such that $G+F$ does not contain
any member of a family ${\mathscr F}$ of forbidden subgraphs. Such an edition set, if any, is called a {\em solution}.
An {\em optimal} edition set $F$ is one with minimum size. We seek for solutions with size at most $k$,
for a nonnegative integer $k$. Clearly, such solutions exist if and only an optimal solution $F$ satisfies $|F| \leq k$.

The notation $-F$ stands for the edition set obtained from
$F$ by replacing each mark $+$ by $-$, and vice versa. $G+F$ and
$G-F$ denote the graphs obtained by applying to $G$ the editions
determined by $F$ and $-F$, respectively. Clearly, $G'=G+F$ if and
only if $G=G'-F$.

The following lemma will be useful:

\begin{lema} \label{orderF}
Let $G$ be a graph, ${\mathscr F}$ a family of forbidden subgraphs, and $F$ a minimum edition set with $|F|=j$ such that $F$ destroys all the members of ${\mathscr F}$ contained in $G$. Then there exists an ordering $\{a_1b_1, \ldots, a_jb_j\}$ of the editions in $F$ such that $a_{i+1}b_{i+1}$ destroys a member of ${\mathscr F}$ contained in $G + F_i$ for every $i \in \{0,\ldots,  j-1\}$, where $F_0=\emptyset$ and $F_i = \{a_1b_1, \ldots, a_ib_i\}$ for $i\geq 1$.
\end{lema}

\noindent
\emph{\textbf{Proof:}} Clearly, the result is valid for edition sets of size $j=1$. Suppose that the result is valid for edition sets of size at most $j-1$, $j>1$. Let $F=\{a_1b_1, \ldots, a_jb_j\}$ be a minimum edition set such that $G'=G+F$ contains no member of ${\mathscr F}$. It is easy to see that there exists at least one edition of $F$ that destroys a member of ${\mathscr F}$ contained in $G$. Without loss of generality, let $a_1b_1$ be this edition. By the induction hypothesis, the result is valid for the edition set $F'= F \backslash \{a_1b_1\}$ when applied to $G + \{a_1b_1\}$. Then, we can obtain the desired ordering of $F$ by appending $a_1b_1$ in the beginning of the ordering of $F'$. We remark that the lemma is also valid for {\it minimal} edition sets. \qed2

\begin{figure}[htb]
\begin{center}
\includegraphics[height=9.33cm,width=13.01cm]{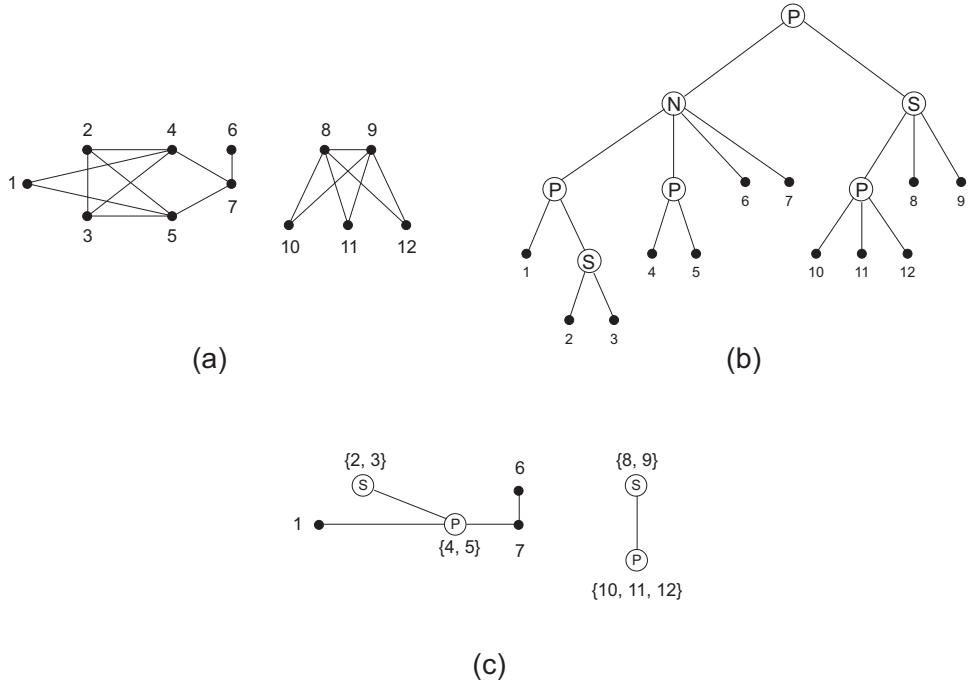}
\caption{\label{grafosQquoc}(a) A graph $G$ \ (b) The modular
decomposition tree $T_{G}$ \ (c) Quotient graph $G_{\mathrm Q}$}
\end{center}
\end{figure}

A subset $M \subseteq V(G)$ is a \emph{module} in $G$ if for all
$x, y \in M$  and $w \in V(G) \backslash M$, $xw \in E(G)$ if
and only if $yw \in E(G)$. A module $M$ is \emph{strong} if,
for every module $M'$, either $M \cap M' = \emptyset$ or one of these
modules is contained in the other. A strong module is
\emph{parallel} when the subgraph induced by its vertices is
disconnected, \emph{series} when the complement of the subgraph
induced by its vertices is disconnected, and \emph{neighborhood}
when both the subgraph induced by its vertices and its complement
are connected. The process of decomposing a graph into strong
modules is called \emph{modular decomposition}. The modular
decomposition of $G$ is represented by a \emph{modular
decomposition tree} $T_{G}$. The nodes of $T_{G}$ correspond to the
strong modules of $G$. The root corresponds to $V(G)$, and the
leaves correspond to the vertices of $G$.  Each internal node of
$T_{G}$ is labeled P (parallel), S (series) or N (neighborhood),
according to the type of the module. The children of every
internal node $M$ of $T_{G}$ are the maximal submodules of $M$.
The modular decomposition tree of a graph is unique up to
isomorphism and can be obtained in linear time \cite{MS94}.
Important references on modular decomposition are \cite{BH83,DGC01,G67,HMC2004,MR84,MS94,RJ2000}.
Figures \ref{grafosQquoc}(a) and \ref{grafosQquoc}(b) show,
respectively, a graph $G$ and its modular decomposition tree
$T_G$.

\subsection{Hardness of $\L${\sc -Cluster Editing}}

To conclude Section 2, we prove that the decision version of $\L${\sc -Cluster Editing} is NP-complete.
We first show that the decision version of $\KL${\sc -Cluster Editing} is NP-complete. Given $G$ and $k$,
$\KL${\sc -Cluster Editing} has answer `yes' if and only if $G$ can be transformed
into a target graph consisting of a disjoint union of $\ell$-cliques by editing at most $k$ edges of $G$.

\begin{lema} \label{NP-hard L-cluster}
Let $\ell \geq 2$. The problem $\KL${\sc -Cluster Editing} is NP-complete.
\end{lema}

\noindent
\emph{\textbf{Proof:}}
Membership in NP is trivial. We remark that the case $\ell=2$ ({\sc Bicluster Editing})
was already shown to be NP-complete by Amit~\cite{TBGEP2004}.
We prove the NP-hardness via a reduction from {\sc Cluster Editing},
which is known to be NP-complete~\cite{SST2004}.

Let $G=(V,E)$ be a graph with $V=\{v_1, \ldots, v_n\}$.
We can assume that $G$ contains no trivial component.
Define $\widetilde{G}$ as follows:

\begin{itemize}

\item $V(\widetilde{G})= \bigcup_{v_i \in V} \{v_i^1, v_i^2, \ldots, v_i^{\ell}\} \ ,$

\item $E(\widetilde{G})= E_1 \cup E_2 \ ,$

\item $E_1 = \{v_i^p v_i^q: \ 1 \leq i \leq n \ , \ 1 \leq p,q \leq \ell \ , \ p\not= q\} \ ,$ and

\item $E_2 = \{ v_i^p v_j^q: \ 1 \leq i,j \leq n \ , \ v_i v_j \in E \ , \ 1 \leq p,q \leq \ell \ , \ p\not=q\}.$

\end{itemize}

In words, for each vertex $v_i \in V$, we construct a clique $Q_i$ with $\ell$ vertices in $\widetilde{G}$,
and for each edge $v_i v_j \in E$, we add all possible edges between $Q_i$ and $Q_j$, except between
vertices with the same superscript. Observe that $\widetilde{G}$ is $\ell$-partite (vertices with the same superscript $p$ form an independent set).

We prove that there exists a solution $F$ of {\sc Cluster Editing} for $G$ with size at most $k$
if and only if there exists a solution $\widetilde{F}$ of $\KL${\sc -Cluster Editing} for $\widetilde{G}$ with size at most $k\ell(\ell-1)$.

Let $F$ be a solution for $G$ with size at most $k$. Define $\widetilde{F}$ as the following edition set for $\widetilde{G}$:

\[\widetilde{F}= \bigcup_{v_iv_j \in F} \{v_i^p v_j^q \ : \ 1 \leq p,q \leq \ell, p \not= q\}.\]

As pointed out before, it is implicit that if $+ v_iv_j \in F$ (resp. $- v_iv_j \in F$) then $+ v_i^p v_j^q \in \widetilde{F}$ (resp. $- v_i^p v_j^q \in \widetilde{F}$) for $1 \leq p,q \leq \ell, p \not= q$. An edge edition $+ v_iv_j$ implies linking $Q_i$ and $Q_j$ in $\widetilde{G}$ by $\ell(\ell-1)$ edges (vertices $v_i^p$ and $v_j^p$ remain unlinked for all $p=1,\ldots,\ell$), and an edge edition $- v_iv_j$ implies disconnecting $Q_i$ and $Q_j$ in $\widetilde{G}$ by removing the $\ell(\ell-1)$ edges between them.

Note that each clique in the cluster graph $G+F$ corresponds to an $\ell$-clique in $\widetilde{G}+\widetilde{F}$. Thus $\widetilde{F}$ is indeed a solution for $\widetilde{G}$, and $|\widetilde{F}| \leq k\ell(\ell-1)$.

Conversely, suppose there exists a minimum solution $F$ for $G$ such that $|F|>k$. Without loss of generality, suppose also $|F|=k+1$. Since $F$ is minimum, by Lemma~\ref{orderF} there exists an ordering $\{v_{i_1}v_{j_1}, v_{i_2}v_{j_2}, \ldots, v_{i_{k+1}}v_{j_{k+1}}\}$ of $F$ such that $v_{i_{h+1}}v_{j_{h+1}}$ destroys a forbidden subgraph in $G + F_h$, $0 \leq h \leq k$, where $F_0=\emptyset$ and $F_h=\{v_{i_1}v_{j_1}, v_{i_2}v_{j_2}, \ldots, v_{i_h}v_{j_h}\}$ for $h\geq 1$. We prove that

\[\widetilde{F}= \bigcup_{v_iv_j \in F} \{v_i^p v_j^q \ : \ 1 \leq p,q \leq \ell, p \not= q\}\]

is a minimum solution for $\widetilde{G}$. Clearly, $\widetilde{F}$ is a solution and $|\widetilde{F}|= (k+1)\ell(\ell -1)$.

\begin{figure}[htb]
\centering
\includegraphics[height=1.9cm,width=5.26cm]{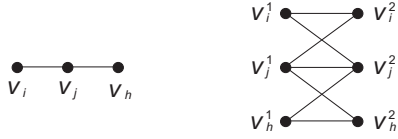}
\caption{\label{casobase} A $P_3$ contained in $G$ and the corresponding induced subgraph in $\widetilde{G}$, for $\ell=2$.}
\end{figure}

For $k=0$, we use induction on $\ell$ to prove the result. Let $v_iv_jv_h$ be the only $P_3$ contained in $G$. When $\ell=2$, there exist three minimum solutions $\widetilde{F}$ for $\widetilde{G}$, each one having size $(k+1)\ell(\ell-1)=2$. Namely, $\{-v_i^1 v_j^2, -v_j^1 v_i^2\}$,  $\{-v_j^1v_h^2, -v_h^1v_j^2\}$ and $\{+v_i^1v_h^2, +v_h^1v_i^2\}$, corresponding to solutions $\{-v_iv_j\}$, $\{-v_j v_h\}$ and $\{+v_i v_h\}$, respectively (see Figure \ref{casobase}).

When $\ell>2$, assume by the induction hypothesis that the result is valid for $\ell -1$. Let $X=\{v_i^\ell \mid v_i \in V(G)\}$ and $\widetilde{H} =  \widetilde{G} - X$. In order to destroy all forbidden subgraphs induced by the subset of vertices $\{v_i^1,\ldots,v_i^{\ell-1}$, $ v_j^1,\ldots,v_j^{\ell-1}, v_h^1,\ldots,v_h^{\ell-1}\}$ in $\widetilde{H}$, $(\ell-1)(\ell-2)$ edge editions are necessary. Besides, there exist three minimum solutions $\widetilde{F_1}$ for $\widetilde{H}$, each of size $(\ell-1)(\ell-2)$, namely $\{-v_i^p v_j^q \mid 1 \leq p,q \leq \ell-1, p\not=q\}$, $\{-v_j^p v_h^q \mid 1 \leq p,q \leq \ell-1, p\not=q\}$ and $\{+v_i^p v_h^q \mid 1 \leq p,q \leq \ell-1, p\not=q\}$. These three cases are analyzed as follows.

\begin{figure}[htb]
\centering
\includegraphics[height=5.94cm,width=10.21cm]{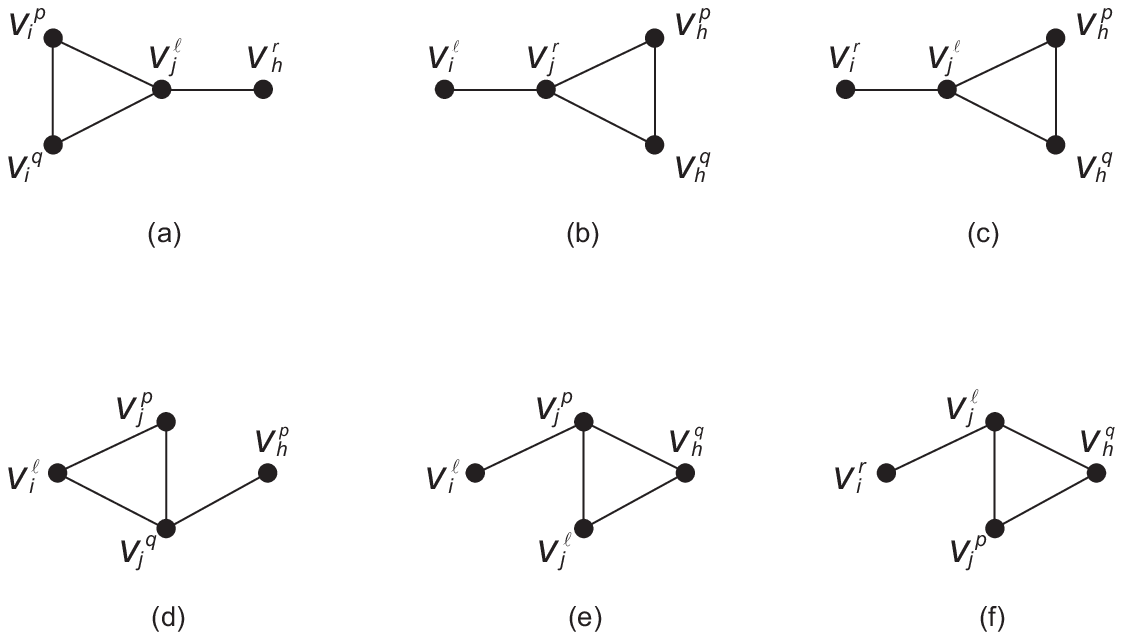}
\caption{\label{caso1}(a) $1 \leq p,q,r \leq \ell-1$, $p\not=q; \;\;$ (b) $1 \leq r \leq \ell-1$, $1 \leq p,q \leq \ell$, $p\not=q\not=r; \;\;$ (c) $1 \leq p,q,r \leq \ell-1$, $p\not=q; \;\;$ (d) $1 \leq p,q \leq \ell-1$, $p\not=q$;  \mbox{(e) $1 \leq p,q \leq \ell-1$, $p\not=q; \;\;$ (f) $1 \leq p,q,r \leq \ell-1$, $p\not=q$}.}
\end{figure}

\noindent
\textbf{Case 1)} $\widetilde{F_1}= \{-v_i^p v_j^q \mid 1 \leq p,q \leq \ell-1$, $p\not=q\}$. Consider $\widetilde{G} + \widetilde{F_1}$. In this graph, it is still necessary to destroy the paws illustrated in Figure \ref{caso1}. The edition subset $\widetilde{F_2}=\{-v_i^\ell v_j^s, -v_i^s v_j^{\ell} \mid 1 \leq s \leq \ell-1\}$ of $\widetilde{F}$ achieves this end. Moreover, the edge editions in $\widetilde{F_2}$ are mandatory, in the sense that excluding one of them from $\widetilde{F}$ leaves a paw undestroyed in $\widetilde{G}$. Since \  $|\widetilde{F_2}|=2(\ell-1)$, we have overall for this case a unique minimum edition set $\widetilde{F} = \widetilde{F_1} \cup \widetilde{F_2} = \{-v_i^p v_j^q \mid 1 \leq p,q \leq \ell$, $p\not=q\}$, whose size is $\ell(\ell-1)$.

\noindent
\textbf{Case 2)} $\widetilde{F_1}= \{-v_j^p v_h^q \mid 1 \leq p,q \leq \ell-1$, $p\not=q\}$. This case is analogous to the previous one.

\begin{figure}[htb]
\centering
\includegraphics[height=2.53cm,width=13.97cm]{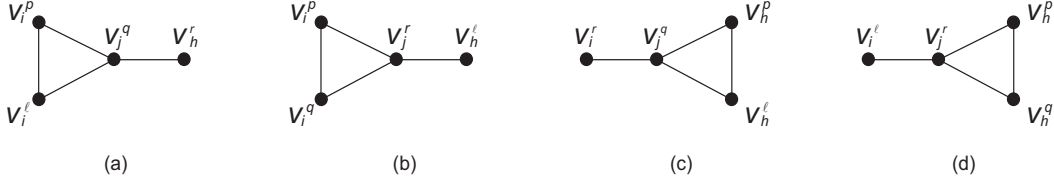}
\caption{\label{caso3}(a) $1 \leq p,q \leq \ell-1$, $p\not=q, r=p$ or $r=\ell;\;\;$
(b) $1 \leq p,q,r \leq \ell-1$, $p\not=q\not=r; \;\;$
(c) $1 \leq p,q \leq \ell-1$, $p\not=q$, $r=p$ or $r=\ell;\;\;$
(d) $1 \leq p,q,r \leq \ell-1$, $p\not=q\not=r$.}
\end{figure}

\noindent
\textbf{Case 3)} $\widetilde{F_1}= \{+v_i^p v_h^q: 1 \leq p,q \leq \ell-1$, $p\not=q\}$. Consider again the graph $\widetilde{G} + \widetilde{F_1}$, and note that several paws still need to be destroyed. Some of them are illustrated in Figure \ref{caso3}.

In order to destroy the forbidden subgraphs of $\widetilde{G} + \widetilde{F_1}$, there exists a unique applicable edition subset of size $2(\ell -1)$, namely
$\widetilde{F_3} = \{+v_i^\ell v_h^s, +v_i^s v_h^{\ell} \mid 1 \leq s \leq \ell-1\}$. Overall, we have for this case a unique minimum edition set
$\widetilde{F} = \widetilde{F_1} \cup \widetilde{F_3} = \{+v_i^p v_h^q \mid 1 \leq p,q \leq \ell$, $p\not=q\}$, whose size is $\ell(\ell-1)$.

As the result is valid for $k=0$, we conclude (using the ordering $\{v_{i_1}v_{j_1}, v_{i_2}v_{j_2}, \ldots, v_{i_{k+1}}v_{j_{k+1}}\}$ of $F$) that the result is valid for any $k>0$. \qed2

\begin{teo}
$\L${\sc -Cluster Editing} is NP-complete.
\end{teo}

\emph{\textbf{Proof:}} Membership in NP is trivial. Let $G$ be an instance of $\KL${\sc -Cluster Editing}.
Recall from the reduction in Lemma~\ref{NP-hard L-cluster} that $\KL${\sc -Cluster Editing} remains NP-complete when restricted to $\ell$-partite graphs.
Hence, assume that $G$ is $\ell$-partite. Define an instance $\widetilde{G}$ for $\L${\sc -Cluster Editing} by setting $\widetilde{G}=G$. We show that there exists a solution for $G$ with size at most $k$ if and only if there exists a solution for $\widetilde{G}$ with size at most $k$. The `only if' part is trivial, since every $\KL$-cluster graph is also an $\L$-cluster graph.
Conversely, let $\widetilde{F}$ be a solution for $\widetilde{G}$, and let $\widetilde{G}_1, \ldots, \widetilde{G}_r$ be the connected components of $\widetilde{G}+\widetilde{F}$.
If these components are all $\ell$-cliques, the result follows. Otherwise, assume that $\widetilde{G}_1$ is a clique but not an $\ell$-clique.
Then $\widetilde{G}_1$ contains at least $\ell+1$ vertices (otherwise it would be $\ell$-partite and thus an $\ell$-clique). Since $\widetilde{G}$ is $\ell$-partite, let $P_1, \ldots, P_\ell$ be the partite sets of $V(\widetilde{G})$,
and consider the subsets $V(\widetilde{G}_1) \cap P_1, \ldots, V(\widetilde{G}_1)\cap P_\ell$.
At least one of these subsets contains more than one vertex. Thus we can construct a new edition set $F$ from $\widetilde{F}$, $|F| < |\widetilde{F}| \leq k$,
by removing from $\widetilde{F}$ the edge additions among vertices of a same subset $V(\widetilde{G}_1) \cap P_i$, for all $1 \leq i \leq \ell$,
and proceeding the same way for all clique components of $\widetilde{G}$ with at least $\ell+1$ vertices. \qed2

\section{Q-quotient graphs}\label{sec3}

In this section we define a special type of graph, namely the Q-\emph{quotient graph}, that allows the establishment of reduction rules for the kernelization algorithm.


\begin{defin}
A partition $\Pi$ of $V(G)$ is the \emph{Q-partition} of $V(G)$ if $\Pi$ satisfies the following conditions:\\
-- if $x \in V(G)$ is a leaf child of a node labelled \emph{N} in $T_G$ then $\{x\}$ is a part of $\Pi$;\\
-- if $x_1, x_2, \ldots, x_j \in V(G)$ are the leaf children of a node labelled \emph{P} or \emph{S} in $T_G$ then $\{x_1, x_2, \ldots, x_j\}$ is a part of $\Pi$.
\end{defin}

A partition $\Pi$ of $V(G)$ such that each part of $\Pi$ is a
module is called \emph{congruence partition}, and the graph whose
vertices are the parts of $\Pi$ and whose edges correspond to the
adjacency relationships involving parts of $\Pi$ is called
\emph{quotient graph} $G/\Pi$.

Clearly, every part of the Q-partition is a strong module in $G$.
Therefore, it is a special type of congruence partition. Since the
modular decomposition tree of a graph is unique, the Q-partition
is also unique.

\begin{defin}
Let $\Pi$ be a partition of $V(G)$. If $\Pi$ is the
\emph{Q}-partition of $V(G)$ then $G/\Pi$ is the \emph{Q-quotient
graph} of $G$, denoted by $G_{\mathrm Q}$.
\end{defin}

A vertex of $G_{\mathrm Q}$ corresponding to a part of $\Pi$ which
contains the children of a node labelled P (resp. S) in $T_G$ is
called P-\emph{vertex} (resp. S-\emph{vertex}); whereas a vertex
corresponding to a singleton $\{x\}$ of $\Pi$ is called
U-\emph{vertex}. We remark that S-vertices can also be seen as
{\em critical cliques}~\cite{G2007}, and P-vertices as {\em
critical independent sets}~\cite{YZ2007}.

Let $M \subseteq V(G)$ be a module corresponding to a P-vertex (or S-vertex).
For simplicity, we write $M$ to stand for both the
module and the P-vertex (S-vertex). Similarly, if a U-vertex is associated with
part $\{x\}$  of $\Pi$ then we write $x$ to stand
for the U-vertex. We also say that a vertex $y \in V(G)$  {\em
belongs} to a P-vertex or an S-vertex $M \in V(G_{\mathrm Q})$ when
$y \in M$.

If $H$ is a Q-quotient graph, denote by ${\mathcal P}(H)$ (resp.
${\mathcal S}(H)$) the set of P-vertices (resp. S-vertices) of
$H$, and by ${\mathcal U}(H)$ the set of U-vertices of $H$.

Figure \ref{grafosQquoc}(c) depicts the graph $G_{\mathrm Q}$ for
the graph $G$ in Figure \ref{grafosQquoc}(a), where P-vertices are
graphically represented by the symbol
\mbox{$\bigcirc$\hspace{-.105in}\scriptsize{P}} \normalsize, and
S-vertices by \mbox{$\bigcirc$\hspace{-.10in}\scriptsize{S}$\;$}
\normalsize.

In the remainder of this work, $F$ denotes an edition set for $G$, and $G'$ denotes the graph $G+F$.

The next lemma presents useful bounds on the sizes of ${\mathcal U}(G'_{\mathrm Q})$,
${\mathcal P}(G'_{\mathrm Q})$, ${\mathcal S}(G'_{\mathrm Q})$ and $V(G'_{\mathrm Q})$
for the case of one edge edition in $G$.

\begin{lema} \label{1edicao-quociente}
Let $F$ be an edition set for $G$, and let $G'=G+F$.
If $|F|=1$ then the following inequalities hold:

\noindent {\em (1)} $|{\mathcal U}(G'_{\mathrm Q})| \leq |{\mathcal U}(G_{\mathrm Q})|+4$.

\noindent {\em (2)} $|{\mathcal P}(G'_{\mathrm Q})| \leq |{\mathcal P}(G_{\mathrm Q})|+2$.

\noindent {\em (3)} $|{\mathcal S}(G'_{\mathrm Q})| \leq |{\mathcal S}(G_{\mathrm Q})|+2$.

\noindent {\em (4)} $|V(G'_{\mathrm Q})|   \leq |V(G_{\mathrm Q})|+2$.
\end{lema}

\noindent \emph{\textbf{Proof:}} Let $x y$ be the edited
edge. The proof is based on the analysis of the local editions
made in $G_{\mathrm Q}$ in order to obtain $G'_{\mathrm Q}$, by
considering the new adjacency relations in $G'$. There are eight
cases, described below.

\medskip

\noindent {\em Case 1}: $x$ and $y$ are U-vertices in $G_{\mathrm
Q}$. In this case $\{x,y\}$ does not form a module in $G$, and
therefore cannot be converted into a P-vertex or an S-vertex in
$G'_{\mathrm Q}$. Since $x, y$ are vertices in $G_{\mathrm Q}$, it
will be useful to regard $F$ also as an edition set of size
one for $G_{\mathrm Q}$, and look at the graph $G_{\mathrm Q} + F$
(which in general is \emph{not} isomorphic to $G'_{\mathrm Q}$).

\medskip

\begin{itemize}

\item[(a)] If there exists a U-vertex $w$ in $G_{\mathrm Q}$ such
that $w$ is nonadjacent to $x$, and $\{x,w\}$ is a module in
$G_{\mathrm Q}+F$, then $\{x,w\}$ is a new P-vertex in
$G'_{\mathrm Q}$.

\item[(b)] If there exists a U-vertex $w$ in $G_{\mathrm Q}$ such
that $w$ is adjacent to $x$, and $\{x,w\}$ is a module in
$G_{\mathrm Q}+F$, then $\{x,w\}$ is a new S-vertex in
$G'_{\mathrm Q}$.

\item[(c)] If there exists a P-vertex $M$ in $G_{\mathrm Q}$ such
that $M$ is nonadjacent to $x$, and $M \cup \{x\}$ is a module in
$G_{\mathrm Q}+F$, then $M \cup \{x\}$ is a new P-vertex in
$G'_{\mathrm Q}$.

\item[(d)] If there exists an S-vertex $M$ in $G_{\mathrm Q}$ such
that $M$ is adjacent to $x$, and $M \cup \{x\}$ is a module in
$G_{\mathrm Q}+F$, then $M \cup \{x\}$ is a new S-vertex in
$G'_{\mathrm Q}$.

\item[(e)] If none of the previous situations (a)-(d) applies to
$x$ then $x$ is still a U-vertex in $G'_{\mathrm Q}$.

\end{itemize}

The same possibilities (a)-(e) are applicable to $y$.

Overall, we have for this case $|{\mathcal U}(G'_{\mathrm Q})| \leq |{\mathcal
U}(G_{\mathrm Q})|$, $|{\mathcal P}(G'_{\mathrm Q})| \leq
|{\mathcal P}(G_{\mathrm Q})|+2$, $|{\mathcal S}(G'_{\mathrm Q})|
\leq |{\mathcal S}(G_{\mathrm Q})|+2$, and $|V(G'_{\mathrm Q})|
\leq |V(G_{\mathrm Q})|$.

\medskip

\noindent {\em Case 2}: $x$ is a U-vertex and $y$ belongs to a
P-vertex $M$ in $G_{\mathrm Q}$.
Write $M=\{y,y_1, y_2, \ldots, y_{\ell}$\}.

If $\ell=1$, we can observe, considering vertex $y_1$, that:
\begin{itemize}
\item[(a)] $y_1$ cannot form a new P-vertex in $G'_{\mathrm Q}$ together with a U-vertex $w$ ($w\not= x$) of $G_{\mathrm Q}$, because $w$ would already belong to $M$ in $G_{\mathrm Q}$. By the same reason, $y_1$ could not be joined to a P-vertex $M' \not= M$ already existing in $G_{\mathrm Q}$.

\item[(b)] $y_1$ cannot form a new S-vertex in $G'_{\mathrm Q}$ together with a U-vertex $w$ ($w\not= x$) of $G_{\mathrm Q}$, because $y$ would be adjacent to $w$ but not to $y_1$ in $G'$. By the same reason, $y_1$ could not be joined to an S-vertex $M'$ already existing in $G_{\mathrm Q}$.

\item[(c)] $y_1$ cannot form a new P-vertex with $x$ in $G'_{\mathrm Q}$ (if they are not adjacent in $G$), because $x$ would already belong to $M$ in $G_{\mathrm Q}$. Besides, $y$ would be adjacent to $x$ but not to $y_1$ in $G'$.

\item[(d)] $y_1$ can form with $x$ a new S-vertex in $G'_{\mathrm Q}$, if $y_1 x \in E(G)$ and $\{y_1,x\}$ is a module in $G'$.
\end{itemize}

Consider vertex $x$. We observe that if $\{y_1, x\}$ is not a new S-vertex then $x$ is still a U-vertex in $G'_{\mathrm Q}$.
With respect to $y$, there are three possibilities: $y$ can be a new U-vertex, $y$ can form a new P-vertex with some U-vertex $w$ of $G_{\mathrm Q}$ ($w \not=x$), or $y$ can be added to a pre-existing P-vertex $M'$ of $G_{\mathrm Q}$.

If $\ell>1$ then $M \backslash \{y\}$ is a P-vertex in $G'_{\mathrm Q}$, since the previous cases (a) and (c) would also be applied (by replacing $y_1$ by $\{y_1, y_2, \ldots, y_{\ell}$\}), and $M \backslash \{y\}$ has at least two nonadjacent vertices (thus cannot be included into an S-vertex). Therefore, in this case $x$ is still a U-vertex in $G'_{\mathrm Q}$. With respect to $y$, the same possibilities of the previous situation are applied.

Overall,  we have for this case
$|{\mathcal U}(G'_{\mathrm Q})| \leq |{\mathcal U}(G_{\mathrm Q})| + 2$,
$|{\mathcal P}(G'_{\mathrm Q})| \leq |{\mathcal P}(G_{\mathrm Q})|+1$,
$|{\mathcal S}(G'_{\mathrm Q})| \leq |{\mathcal S}(G_{\mathrm Q})|+1$, and
$|V(G'_{\mathrm Q})| \leq |V(G_{\mathrm Q})| + 1$.

\medskip

\noindent {\em Case 3}: $x$ is a U-vertex and $y$ belongs to an
S-vertex $M$ in $G_{\mathrm Q}$. Write $M=\{y,y_1, y_2, \ldots, y_{\ell}$\}.
%
%

If $\ell=1$, we can observe, considering vertex $y_1$, that:
\begin{itemize}
\item[(a)] $y_1$ cannot form a new S-vertex in $G'_{\mathrm Q}$ together with a U-vertex $w$ ($w\not= x$) of $G_{\mathrm Q}$, because $w$ would already belong to $M$ in $G_{\mathrm Q}$. By the same reason, $y_1$ could not be joined to an S-vertex $M' \not= M$ already existing in $G_{\mathrm Q}$.

\item[(b)] $y_1$ cannot form a new P-vertex in $G'_{\mathrm Q}$ together with a U-vertex $w$ ($w\not= x$) of $G_{\mathrm Q}$, because $y$ would be adjacent to $y_1$ but not to $w$ in $G'$. By the same reason, $y_1$ could not be joined to a P-vertex $M'$ already existing in $G_{\mathrm Q}$.

\item[(c)] $y_1$ cannot form a new S-vertex with $x$ in $G'_{\mathrm Q}$ (if they are adjacent in $G$), because $x$ would already belong to $M$ in $G_{\mathrm Q}$. Besides, $y$ would be adjacent to $y_1$ but not to $x$ in $G'$.

\item[(d)] $y_1$ can form with $x$ a new P-vertex in $G'_{\mathrm Q}$, if $y_1x \notin E(G)$ and $\{y_1,x\}$ is a module in $G'$.
\end{itemize}

Considering vertex $x$, we observe that if $\{y_1, x\}$ is not a new P-vertex then $x$ is still a U-vertex in $G'_{\mathrm Q}$.
With respect to $y$, there are three possibilities: $y$ can be a new U-vertex, $y$ can form a new S-vertex with some U-vertex $w$ of $G_{\mathrm Q}$ ($w \not=x$), or $y$ can be added to a pre-existing S-vertex $M'$ of $G_{\mathrm Q}$.

If $\ell>1$ then $M \backslash \{y\}$ is an S-vertex in $G'_{\mathrm Q}$, since the previous cases (a) and (c) would also be applied (by replacing $y_1$ by $\{y_1, y_2, \ldots, y_{\ell}$\}), and $M \backslash \{y\}$ has at least two adjacent vertices (thus cannot be included into a P-vertex). Therefore, in this case $x$ is still a U-vertex in $G'_{\mathrm Q}$. With respect to $y$, the same possibilities  of the previous situation are applied.

Overall, we have for this case
$|{\mathcal U}(G'_{\mathrm Q})| \leq |{\mathcal U}(G_{\mathrm Q})| + 2$,
$|{\mathcal P}(G'_{\mathrm Q})| \leq |{\mathcal P}(G_{\mathrm Q})|+1$,
$|{\mathcal S}(G'_{\mathrm Q})| \leq |{\mathcal S}(G_{\mathrm Q})|+1$, and
$|V(G'_{\mathrm Q})| \leq |V(G_{\mathrm Q})| + 1$.

\medskip

\noindent {\em Case 4}: $x$ and $y$ belong to distinct P-vertices
$M$ and $M'$ in $G_{\mathrm Q}$, respectively.
Write $M=\{x,x_1,\ldots,x_{\ell}\}$ and $M'=\{y,y_1,\ldots,y_r\}$. Then $x$ and $y$ are two new U-vertices in $G'_{\mathrm Q}$. If $\ell=1$ and $r=1$, $x_1$ and $y_1$ are also two new U-vertices in $G'_{\mathrm Q}$. If $\ell=1$ and $r>1$, $x_1$ is a new U-vertex and $M' \backslash \{y\}$ is a P-vertex in $G'_{\mathrm Q}$. The situation $\ell>1$ and $r=1$ is similar to the previous one. Finally, if $\ell,r >1$ then $M \backslash \{x\}$ and $M' \backslash \{y\}$ are P-vertices in $G'_{\mathrm Q}$.

Overall,
we have for this case
$|{\mathcal U}(G'_{\mathrm Q})| \leq |{\mathcal U}(G_{\mathrm Q})|+4$,
$|{\mathcal P}(G'_{\mathrm Q})| \leq |{\mathcal P}(G_{\mathrm Q})|$,
$|{\mathcal S}(G'_{\mathrm Q})| = |{\mathcal S}(G_{\mathrm Q})|$, and
$|V(G'_{\mathrm Q})|=|V(G_{\mathrm Q})|+2.$

\medskip

\noindent {\em Case 5}: $x$ and $y$ belong to the same P-vertex $M$ in $G_{\mathrm Q}$.
Write $M=\{x,y\} \cup W$. Then $xy$ is an added edge. The vertex $x$ cannot form a new P-vertex $M'$, because $y$ would be adjacent to $x$ but not to $M' \backslash x$ (the same applies to $y$). Vertex $x$ cannot either form a new S-vertex $M'$ with a U-vertex $w$ (or with another \makebox{S-vertex}), because $W$ would be adjacent to $M'\backslash x$, but not to $x$ (the same applies to $y$). Since $\{x,y\}$ is still a module in $G'$, $\{x,y\}$  forms a new S-vertex in $G'_{\mathrm Q}$. If $|W|=1$ then $W$ is a new U-vertex, and if $|W|>1$ then $W$ is a P-vertex in $G'_{\mathrm Q}$.

Overall,
we have for this case
$|{\mathcal U}(G'_{\mathrm Q})| \leq |{\mathcal U}(G_{\mathrm Q})|+1$,
$|{\mathcal P}(G'_{\mathrm Q})| \leq |{\mathcal P}(G_{\mathrm Q})|$,
$|{\mathcal S}(G'_{\mathrm Q})| = |{\mathcal S}(G_{\mathrm Q})|+1$, and
$|V(G'_{\mathrm Q})|=|V(G_{\mathrm Q})|+1.$

\medskip

\noindent {\em Case 6}: $x$ and $y$ belong to distinct S-vertices
$M$ and $M'$ in $G_{\mathrm Q}$, respectively.
Write $M=\{x,x_1,\ldots,x_{\ell}\}$ and $M'=\{y,y_1,\ldots,y_r\}$. Then $x$ and $y$ are two new U-vertices in $G'_{\mathrm Q}$. If $\ell=1$ and $r=1$, $x_1$ and $y_1$ are also two new U-vertices in $G'_{\mathrm Q}$. If $\ell=1$ and $r>1$, $x_1$ is a new U-vertex and $M' \backslash \{y\}$ is an S-vertex in $G'_{\mathrm Q}$. The situation $\ell>1$ and $r=1$ is similar to the previous one. Finally, if $\ell,r >1$ then $M \backslash \{x\}$ and $M' \backslash \{y\}$ are S-vertices in $G'_{\mathrm Q}$.

Overall,
we have for this case
$|{\mathcal U}(G'_{\mathrm Q})| \leq |{\mathcal U}(G_{\mathrm Q})|+4$,
$|{\mathcal P}(G'_{\mathrm Q})| = |{\mathcal P}(G_{\mathrm Q})|$,
$|{\mathcal S}(G'_{\mathrm Q})| \leq |{\mathcal S}(G_{\mathrm Q})|$, and
$|V(G'_{\mathrm Q})|=|V(G_{\mathrm Q})|+2.$

\medskip

\noindent {\em Case 7}: $x$ and $y$ belong to the same S-vertex $M$ in $G_{\mathrm Q}$.
Write $M=\{x,y\} \cup W$. Then $xy$ is a removed edge. Vertex $x$ cannot form a new S-vertex $M'$, because $y$ would be adjacent to $M' \backslash x$ but not to $x$ (the same applies to $y$). Vertex $x$ cannot either form a new P-vertex $M'$ with a U-vertex $w$ (or with another \makebox{P-vertex}), because $W$ would be adjacent to $x$  but not to $M'\backslash x$ (the same applies to $y$). Since $\{x,y\}$ is still a module in $G'$, $\{x,y\}$  forms a new P-vertex in $G'_{\mathrm Q}$. If $|W|=1$ then $W$ is a new U-vertex, and if $|W|>1$ then $W$ is an S-vertex in $G'_{\mathrm Q}$.

Overall,
we have for this case
$|{\mathcal U}(G'_{\mathrm Q})| \leq |{\mathcal U}(G_{\mathrm Q})|+1$,
$|{\mathcal P}(G'_{\mathrm Q})| = |{\mathcal P}(G_{\mathrm Q})|+1$,
$|{\mathcal S}(G'_{\mathrm Q})| \leq |{\mathcal S}(G_{\mathrm Q})|$, and
$|V(G'_{\mathrm Q})|=|V(G_{\mathrm Q})|+1.$

\medskip

\noindent {\em Case 8}: $x$ belongs to a P-vertex $M$ and $y$
belongs to an S-vertex $M'$ in  $G_{\mathrm Q}$.
Write $M=\{x,x_1,\ldots,x_{\ell}\}$ and $M'=\{y,y_1,\ldots,y_r\}$. Consider vertex $x$. We observe that:

\begin{itemize}
\item[(a)] $x$ cannot form with $y$ a new P-vertex neither a new S-vertex in $G'_{\mathrm Q}$, because $M\backslash \{x\}$ or $M'\backslash\{y\}$ would be adjacent to one vertex of $\{x,y\}$, but not to the other;

\item[(b)] if $r=1$, $x$ is not adjacent to $y_1$ and $\{x,y_1\}$ is a module in $G'$ (thus $\{x,y_1\}$ forms a new P-vertex in $G'_{\mathrm Q}$);

\item[(c)] $x$ cannot be joined to a U-vertex neither to an S-vertex (or a P-vertex) already existing in $G_{\mathrm Q}$ to form a new S-vertex (P-vertex) in $G'_{\mathrm Q}$.
\end{itemize}

Consider now vertex $y$. The following facts hold:
\begin{itemize}
\item[(a)] if $\ell=1$, $y$ is adjacent to $x_1$ and $\{y,x_1\}$ is a module in $G'$ (thus $\{y,x_1\}$ forms a new S-vertex in $G'_{\mathrm Q}$);

\item[(b)] $y$ cannot be joined to a U-vertex neither to an S-vertex (or a P-vertex) already existing in $G_{\mathrm Q}$  to form a new S-vertex (P-vertex) in $G'_{\mathrm Q}$.
\end{itemize}

If $\ell=1$, $x_1$ can also be a new U-vertex; otherwise, $M\backslash \{x\}$ is a P-vertex in $G'_{\mathrm Q}$. If $r=1$, $y_1$ can also be a new U-vertex; otherwise, $M'\backslash \{y\}$ is an S-vertex in $G'_{\mathrm Q}$.
Overall,
we have for this case
$|{\mathcal U}(G'_{\mathrm Q})| \leq |{\mathcal U}(G_{\mathrm Q})|+4$,
$|{\mathcal P}(G'_{\mathrm Q})| \leq |{\mathcal P}(G_{\mathrm Q})|+1$,
$|{\mathcal S}(G'_{\mathrm Q})| \leq |{\mathcal S}(G_{\mathrm Q})|+1$, and
$|V(G'_{\mathrm Q})|\leq|V(G_{\mathrm Q})|+2.$

\medskip

All the cases have been analyzed, thus the lemma follows. \hfill \framebox[.09in]

\bigskip

\section{Building the Problem Kernel} \label{sec4}

Clearly, connected components of the input graph $G$ that are
already cliques or $\ell$-cliques can be omitted from
consideration.

\begin{figure}[htb]
\centering
\includegraphics[height=2.27cm,width=8.68cm]{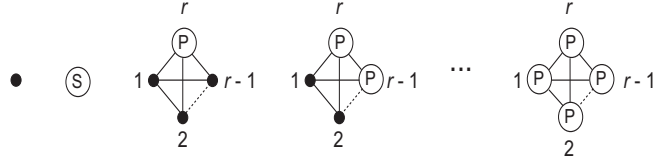}
\caption{\label{q-quocCliqueBiclique}Possible graphical
representations of a connected component of $G'_{\mathrm Q}$,
where $2 \leq r \leq \ell$.}
\end{figure}

If $G' = G+F$, $|F|\leq k$, is an $\L$-cluster graph then $G'$ contains at
most $2k$ connected components. In graph $G'_{\mathrm Q}$, each of them
can have one of the graphical representations illustrated in
Figure \ref{q-quocCliqueBiclique}.

Lemma \ref{kernel1mod} presents bounds on the sizes of ${\mathcal P}(G_{\mathrm Q})$,
${\mathcal S}(G_{\mathrm Q})$ and $V(G_{\mathrm Q})$ when $|F|=1$ and $G'$
is an $\L$-cluster graph.

\begin{lema} \label{kernel1mod}
If $G$ contains no clique or $\ell$-clique component and there
exists an edition set $F$ for $G$ such that $|F|=1$ and
$G'=G+F$ is an $\L$-cluster graph then $|V(G_{\mathrm Q})|
\leq 2 \ell + 2$, $|{\mathcal P}(G_{\mathrm Q})| \leq 2 \ell$ and
$|{\mathcal S}(G_{\mathrm Q})| \leq 2$.
\end{lema}

\emph{\textbf{Proof:}} Since $G = G'-F$, we can apply to $G'$ the inverse
edition in $F$ in order to obtain $G$. Graph $G'$
contains at most $2$ connected components. All the cases in the
proof of Lemma \ref{1edicao-quociente} can then be applied.
The proof follows by analyzing the worst case for each case.

\medskip

\noindent {\em Case 1}: $x$ and $y$ are U-vertices in $G'_{\mathrm Q}$. In the worst case (maximizing the total number of vertices of $G'_{\mathrm Q}$), $x$ and $y$ belong to distinct $\ell$-cliques. By applying the limits established by this case in the proof of Lemma \ref{1edicao-quociente}, we have  $|V(G_{\mathrm Q})| \leq 2\ell $,  $|{\mathcal P}(G_{\mathrm Q})| \leq 2\ell$  (in the worst case each $\ell$-clique contains $\ell -1$ P-vertices) and $|{\mathcal S}(G_{\mathrm Q})| \leq 2$.

\medskip

\noindent {\em Case 2}: $x$ is a U-vertex and $y$ belongs to a P-vertex in $G'_{\mathrm Q}$. In the worst case, $x$ and $y$ belong to distinct $\ell$-cliques. By applying again the limits established by this case in Lemma \ref{1edicao-quociente}, we have $|V(G_{\mathrm Q})| \leq 2\ell +1$, $|{\mathcal P}(G_{\mathrm Q})| \leq 2\ell$ (in the worst case the $\ell$-clique of $y$ contains $\ell$ P-vertices and the $\ell$-clique of $x$ contains $\ell -1$ P-vertices) and $|{\mathcal S}(G_{\mathrm Q})| \leq 1$.

\medskip

\noindent {\em Case 3}: $x$ is a U-vertex and $y$ belongs to an S-vertex in $G'_{\mathrm Q}$. Clearly, $x$ and $y$ belong to distinct connected components. In the worst case, we have $|V(G_{\mathrm Q})| \leq \ell +2$, $|{\mathcal P}(G_{\mathrm Q})| \leq \ell$ and $|{\mathcal S}(G_{\mathrm Q})| \leq 2$.

\medskip

\noindent {\em Case 4}: $x$ and $y$ belong to distinct P-vertices in $G'_{\mathrm Q}$. In the worst case, we have $|V(G_{\mathrm Q})| = 2\ell +2$, $|{\mathcal P}(G_{\mathrm Q})| \leq 2\ell$ and $|{\mathcal S}(G_{\mathrm Q})| =0$.

\medskip

\noindent {\em Case 5}: $x$ and $y$ belong to the same P-vertex in $G'_{\mathrm Q}$. In the worst case, $|V(G_{\mathrm Q})| = \ell +1$, $|{\mathcal P}(G_{\mathrm Q})| \leq \ell$ and $|{\mathcal S}(G_{\mathrm Q})| =1$.

\medskip

\noindent {\em Case 6}: $x$ and $y$ belong to distinct S-vertices in $G'_{\mathrm Q}$. In the worst case, $|V(G_{\mathrm Q})| =4$,   $|{\mathcal P}(G_{\mathrm Q})| =0$ and $|{\mathcal S}(G_{\mathrm Q})| \leq 2$.

\medskip

\noindent {\em Case 7}: $x$ and $y$ belong to the same S-vertex in $G'_{\mathrm Q}$. In the worst case, $|V(G_{\mathrm Q})| =2$, $|{\mathcal P}(G_{\mathrm Q})| =1$ and $|{\mathcal S}(G_{\mathrm Q})| \leq 1$.

\medskip

\noindent {\em Case 8}: $x$ belongs to a P-vertex and $y$ belongs to an S-vertex in $G'_{\mathrm Q}$. In the worst case, $|V(G_{\mathrm Q})| \leq  \ell+3$, $|{\mathcal P}(G_{\mathrm Q})| \leq \ell +1$ and $|{\mathcal S}(G_{\mathrm Q})| \leq 2$.
\hfill \framebox[.09in] \\ \\

The next theorem generalizes the previous lemma.

\begin{teo} \label{sizeG_Q}
If $G$ contains no clique or $\ell$-clique component and there
exists an edition set $F$ for $G$ such that $|F|=k$ and
$G'=G+F$ is an $\L$-cluster graph then $|V(G_{\mathrm Q})|
\leq (2\ell +2)k$, $|{\mathcal P}(G_{\mathrm Q})| \leq 2\ell k$
and $|{\mathcal S}(G_{\mathrm Q})| \leq 2k$.
\end{teo}

\emph{\textbf{Proof:}} Since $G = G'-F$, we can apply to $G'$ the inverse
editions in $F$ in order to obtain $G$. The proof is by induction on $k$.
The basis of the induction is given by Lemma \ref{kernel1mod}.

Let $F^{-}$ be a subset of $F$ such that $|F^{-}|=|F|-1$, and let
$G^{-}=G'- F^{-}$. By the induction hypothesis, the result is
valid for $F^{-}$. Hence, the subgraph of ${(G^{-})}_{\mathrm Q}$
induced by components which are not cliques or $\ell$-cliques
contains at most $(k-1)(2\ell+2)$ vertices, among which at most
$2\ell(k-1)$ are P-vertices and at most $2(k-1)$ are S-vertices.
Since $G'$ can contain $2k$ components, ${(G^{-})}_{\mathrm Q}$
can possibly contain some other components which are cliques or
$\ell$-cliques.

Let $\alpha$ be the edge edition such that $F= F^- \cup
\{\alpha\}$. Then $G= G^- - \{\alpha\}$. Let $x,y$ be the vertices
of $\alpha$. All the cases of the proof of Lemma
\ref{1edicao-quociente} can be applied, by considering all the situations
for vertices $x$ and $y$. Again, the proof follows by analyzing the worst case for each of them. \hfill \framebox[.09in]

\subsection{Splitting P-vertices and S-vertices}

\begin{figure}[htb]
\centering
\includegraphics[height=4.43cm,width=14.25cm]{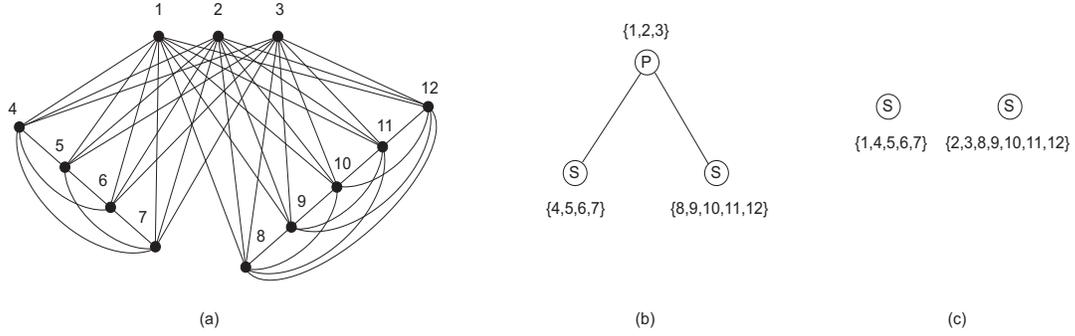}
\caption{\label{splitPvertex}(a) Input graph G; (b) Graph $G_{\mathrm Q}$ (c) Graph $G'_{\mathrm Q}$
obtained from $G+F$, where $F=\{-\;1\;8$, $-\;1\;9$, $-\;1\;10$, $-\;1\;11$, $-\;1\;12$,
$-\;2\;4$, $-\;2\;5$, $-\;2\;6$, $-\;2\;7$, $-\;3\;4$, $-\;3\;5$, $-\;3\;6$, $-\;3\;7$, $+\;2\;3\}$
is an optimal solution for {\sc Mixed Cluster Editing} ($\ell=2$).}
\end{figure}

When there exists an optimal solution with size $k$ of $\L${\sc -Cluster Editing}
such that no P-vertex or S-vertex $M$ of $G_{\mathrm Q}$ is
split into distinct vertices of $G'_{\mathrm Q}$,
the size of $M$ is bounded by $k+1$ \cite{DSPS2006,DSPS2009}.
However, such an optimal solution may not exist.
For instance, let $\ell=2$ and consider the graph $G$ depicted in Figure \ref{splitPvertex}. We have
three optimal solutions for $\L${\sc -Cluster Editing} in this case; all of
them split the P-vertex $M=\{1,2,3\}$ of $G_{\mathrm Q}$ into two S-vertices
of $G'_{\mathrm Q}$. One of the solutions is illustrated in Figure \ref{splitPvertex}(c).

In order to build a problem kernel for $\L${\sc -Cluster Editing}($k$),
we will obtain a bound on the size of P-vertices and S-vertices of
$G_{\mathrm Q}$, by analyzing all possible cases
in which a P-vertex or an S-vertex of $G_{\mathrm Q}$ is split into distinct vertices
of $G'_{\mathrm Q}$ in an optimal solution.

In this subsection, we analyze all possible cases in which two distinct vertices of $G'_{\mathrm Q}$ contain
vertices of a same P-vertex or S-vertex of $G_{\mathrm Q}$.
When a contradiction arises, the assumed split
does not occur in any optimal solution and can be disregarded.
In the next subsection, the analysis is generalized for several
vertices of $G'_{\mathrm Q}$ containing vertices of the same P-vertex or
S-vertex of $G_{\mathrm Q}$. Since at most $k$ edge editions are
allowed, a bound on the size of a P-vertex or an S-vertex of $G_{\mathrm Q}$
that is split into distinct vertices of $G'_{\mathrm Q}$ in an optimal solution can be derived.


Let $M$ be a P-vertex (or S-vertex) of $G_{\mathrm Q}$ whose
vertices are split into distinct vertices of $G'_{\mathrm Q}$, in an
optimal solution $F$. Let $A$ and $B$ be two vertices of
$G'_{\mathrm Q}$ that contain vertices of $M$. Let $X = M \cap A$
and $Y = M \cap B$. We denote by $F_X$ the edition subset of
$F$ containing the editions of the form $ab$ such that $a \in
X$ and $b \notin X \cup Y$. Similarly, $F_Y$ denotes the edition subset of $F$ containing the editions of the form $ab$
such that $a \in Y$ and $b \notin X \cup Y$. Let $x = |X|$ and $y
= |Y|$.

Since $X$ is a module in $G$ (because $X$ is contained in a vertex of $G_{\mathrm Q}$)
and is still a module in $G'$ (because $X$ is contained in a vertex of $G'_{\mathrm Q}$),
the edge editions in $F$ with only one endpoint in $X$ are
replicated for each vertex of $X$, considering the other endpoint and the edition type.
That is, if there exists in $F_X$ an edition $ab$ such that
$a \in X$ then there exists the edition $wb$ in $F_X$,
for all $w \in X$. Hence, we have $|F_X|/x$ editions in
$F_X$ for each vertex of $X$.

The same argument above applies to $Y$.

We now analyze the following cases.

\textbf{Case 1. $A$ and $B$ belong to the same connected
component.}\ \ Then $A$ (or $B$) can be a U-vertex or
P-vertex of an $\ell$-clique in $G'$.

\textbf{Case 1.1.} $M$ is a P-vertex. Then the total number of
editions in $F$ involving vertices of $X \cup Y$ is
$|F_X|+|F_Y|+xy$.

$\bullet$ If $|F_X| \geq \frac{x|F_Y|}{y} + x^2$ then $|F_X|>
\frac{x|F_Y|}{y}-xy$. We have $\frac{(x+y)|F_Y|}{y}<|F_X|+
|F_Y|+xy$. We obtain a smaller edition set if the vertices of
$X$ belong to $B$, and a contradiction follows.

$\bullet$ If $|F_X| < \frac{x|F_Y|}{y} + x^2$ then
$\frac{(x+y)|F_X|}{x}<|F_X|+ |F_Y|+xy$. We obtain a smaller edition set if the vertices of $Y$ belong to $A$, and a
contradiction follows.

\textbf{Case 1.2.} $M$ is an S-vertex. Then the total number of
editions in $F$ involving vertices of $X \cup Y$ is $|F_X|+|F_Y|+
\frac{x(x-1)+y(y-1)}{2}$.

$\bullet$ If $|F_X| \leq \frac{x|F_Y|}{y} - x^2$ then
$\frac{(x+y)|F_X|}{x}+ \frac{(x+y)(x+y-1)}{2}\leq |F_X| + |F_Y| +
\frac{x(x-1)+y(y-1)}{2}$. We can obtain an edition set $F'$,
$|F'|\leq|F|$, if the vertices of $Y$ belong to $A$.

$\bullet$ If $|F_X|\geq \frac{x|F_Y|}{y}+ xy$ then
$\frac{(x+y)|F_Y|}{y}+ \frac{(x+y)(x+y-1)}{2}\leq |F_X| + |F_Y| +
\frac{x(x-1)+y(y-1)}{2}$. We can obtain an edition set $F'$,
$|F'|\leq|F|$, if the vertices of $X$ belong to $B$.

$\bullet$ If $\frac{x|F_Y|}{y}-x^2 < |F_X|< \frac{x|F_Y|}{y}+ xy$ then there is no contradiction,
and we cannot construct an edition set $F'$ with $|F'| \leq |F|$ by applying to all vertices
of $X \cup Y$ the same edge editions.

\textbf{Case 2. $A$ and $B$ belong to distinct connected
components.}

\textbf{Case 2.1.} $A$ and $B$ are clique components of $G'$ (each of them is a U-vertex or an S-vertex in $G'_{\mathrm Q}$).

\textbf{2.1.1.} $M$ is an S-vertex. The total number of editions in
$F$ involving vertices of $X \cup Y$ is $|F_X|+|F_Y|+ xy$.

$\bullet$ If $|F_X|\geq \frac{x |F_Y|}{y} + x^2$ then $|F_X|>
\frac{x |F_Y|}{y}- xy$. Thus $\frac{(x+y)|F_Y|}{y}< |F_X| + |F_Y|
+ xy$.  A smaller edition set is obtained if the vertices of
$X$ belong to $B$, and a contradiction follows.

$\bullet$ If $|F_X|< \frac{x |F_Y|}{y} + x^2$ then
$\frac{(x+y)|F_X|}{x}< |F_X| + |F_Y| + xy$. A smaller edition set is obtained if the vertices of $Y$ belong to $A$, and a
contradiction follows.

\textbf{2.1.2.} $M$ is a P-vertex. The total number of editions in
$F$ involving vertices of $X \cup Y$ is
$|F_X|+|F_Y|+\frac{x(x-1)\, +\, y(y-1)}{2}$.

$\bullet$ If $|F_X| \geq \frac{x |F_Y|}{y}+ xy$ then
$\frac{(x+y)|F_Y|}{y}+ \frac{(x+y)(x+y-1)}{2} \leq |F_X| + |F_Y| +
\frac{x(x-1)\, +\, y(y-1)}{2}$. We can obtain $F'$ with $|F'|\leq
|F|$ if the vertices of $X$ belong to $B$.

$\bullet$ If $|F_X|\leq \frac{x |F_Y|}{y}- x^2$ then
$\frac{(x+y)|F_X|}{x}+ \frac{(x+y)(x+y-1)}{2} \leq |F_X| + |F_Y| +
\frac{x(x-1)\, +\, y(y-1)}{2}$. We can obtain $F'$ with $|F'|\leq
|F|$ if the vertices of $Y$ belong to $A$.

$\bullet$ If $ \frac{x |F_Y|}{y}- x^2 < |F_X| < \frac{x |F_Y|}{y}+
xy$ then there is no contradiction, and we cannot construct $F'$
with $|F'| \leq |F|$ by applying to all vertices of $X \cup Y$ the
same edge editions.

\textbf{Case 2.2.} $A$ is a clique component of size one
(therefore a U-vertex in $G'_{\mathrm Q}$), and $B$ is contained
in an $\ell$-clique component ($B$ is a U-vertex or a P-vertex in
$G'_{\mathrm Q}$).

\textbf{2.2.1.} $M$ is an S-vertex. The total number of editions
in $F$ involving vertices of $X \cup Y$ is $|F_X|+|F_Y|+
\frac{y(y+1)}{2}$.

$\bullet$ If $|F_X| > \frac{|F_Y|}{y}$ then $\frac{(y+1)|F_Y|}{y}
+ \frac{y(y+1)}{2} \leq |F_X|+|F_Y|+ \frac{y(y+1)}{2}$. We can
obtain a smaller edition set by applying to $X$ the same edge
editions applied to $Y$ by $F_Y$ (instead of applying $F_X$), and
a contradiction follows.

$\bullet$ If  $|F_X| \leq \frac{|F_Y|}{y}$ then $|F_X| <
\frac{|F_Y|}{y}+ \frac{(y+1)}{2}$. Thus $(y+1)|F_X| < |F_X|+|F_Y|+
\frac{y(y+1)}{2}$. We can obtain a smaller edition set by
applying to $Y$ the same edge editions applied to $X$ by $F_X$,
and a contradiction follows.

\textbf{2.2.2.} $M$ is a P-vertex. The total number of editions in
$F$ involving vertices of $X \cup Y$ is $|F_X|+|F_Y|$.

$\bullet$ If $|F_X| > \frac{|F_Y|}{y}$ then $\frac{(y+1)|F_Y|}{y}
< |F_X|+|F_Y|$. We can obtain a smaller edition set by
applying to $X$ the same edge editions applied to $Y$ by $F_Y$,
and a contradiction follows.

$\bullet$ If  $|F_X| \leq \frac{|F_Y|}{y}$ then $(y+1)|F_X| \leq
|F_X|+|F_Y|$. We can obtain an edition set $F'$ with
$|F'|\leq|F|$ by applying $F_X$ to each vertex of $Y$.

\textbf{Case 2.3.} $A$ is a clique component of size at least two
(an S-vertex in $G'_{\mathrm Q}$), and $B$ is contained in an
$\ell$-clique component ($B$ is a U-vertex or a P-vertex in
$G'_{\mathrm Q}$).

\textbf{2.3.1.} $M$ is a P-vertex. The total number of editions in
$F$ involving vertices of $X \cup Y$ is
$|F_X|+|F_Y|+\frac{x(x-1)}{2}$.

$\bullet$ If $|F_X| \geq \frac{x|F_Y|}{y}- \frac{x(x-1)}{2}$ then
$\frac{(x+y)|F_Y|}{y} \leq |F_X|+|F_Y|+ \frac{x(x-1)}{2}$. We can
obtain $F'$ with $|F'|\leq |F|$ if the vertices of $X$ belong to
$B$.

$\bullet$ If $|F_X|\leq \frac{x|F_Y|}{y} - \frac{x(2x +y-1)}{2}$ then $\frac{(x+y)|F_X|}{x} + \frac{(x+y)(x+y-1)}{2} \leq
|F_X|+|F_Y|+ \frac{x(x-1)}{2}$. We can obtain $F'$ with $|F'|\leq
|F|$ if the vertices of $Y$ belong to $A$.

$\bullet$ If $\frac{x|F_Y|}{y}- \frac{x(2x+y-1)}{2} < |F_X|<
\frac{x|F_Y|}{y} - \frac{x(x-1)}{2}$ then there is no
contradiction, and we cannot construct $F'$ with $|F'| \leq |F|$ by
applying to all vertices of $X \cup Y$ the same edge editions.

\textbf{2.3.2.} $M$ is an S-vertex. The total number of editions
in $F$ involving vertices of $X \cup Y$ is
$|F_X|+|F_Y|+xy+\frac{y(y-1)}{2}$.

$\bullet$ If $|F_X| > \frac{x|F_Y|}{y}+ \frac{x(x-1)}{2}$ then
$\frac{(x+y)|F_Y|}{y} + \frac{(x+y)(x+y-1)}{2}< |F_X| + |F_Y| + xy
+  \frac{y(y-1)}{2}$. A smaller edition set is obtained if
the vertices of $X$ belong to $B$, and a contradiction follows.

$\bullet$ If $|F_X| \leq \frac{x|F_Y|}{y}+ \frac{x(x-1)}{2}$ then
$|F_X| < \frac{x|F_Y|}{y}+ \frac{x(2x+y-1)}{2}$. Thus
$\frac{(x+y)|F_X|}{x}< |F_X| + |F_Y| + xy +  \frac{y(y-1)}{2}$. A
smaller edition set is obtained if the vertices of $Y$ belong
to $A$, and a contradiction follows.

\textbf{Case 2.4.} $A$ and $B$ are contained in $\ell$-clique
components of $G'$ (each of them is a U-vertex or P-vertex in
$G'_{\mathrm Q}$). In this case, $M$ can be a P-vertex or an
S-vertex, since in both cases $X \cup Y$ is an independent
set in $G'$.

$\bullet$ If $|F_X| > \frac{x|F_Y|}{y}$, we can obtain a smaller
edition set by applying to $X$ the same edge editions applied
to $Y$ by $F_Y$, and a contradiction follows.

$\bullet$ If  $|F_X| \leq \frac{x|F_Y|}{y}$, we can obtain an edition set $F'$ with $|F'|\leq|F|$ by applying to $Y$ the same
edge editions applied to $X$ by $F_X$.

\subsection{Determining the kernel's size}

By analyzing all the cases previously described, we observe that it is
often possible to replace an optimal solution containing the
split of a P-vertex or S-vertex by another optimal solution in
which this split does not occur. However, there exist some
unavoidable splits, described below. We analyze these
cases in order to bound the size of P-vertices and S-vertices in
the problem kernel.

\emph{\textbf{Splitting an S-vertex.}} An S-vertex can be
split into distinct vertices of the same $\ell$-clique of
$G'_{\mathrm Q}$. If an S-vertex $M$ contains more than $\ell+k$
vertices then, given a solution $F$ such that $|F|\leq k$, no
vertex of $M$ is an endpoint of an edge edition in $F$, since each
edge edition can decrease the chromatic number of a clique by at
most one and $M$ induces an $\ell$-clique in $G'$.

\emph{\textbf{Splitting a P-vertex.}} There are two cases for
the split of a P-vertex $M$:

1) Only cliques in $G'$ contain vertices of $M$. Since $G$
contains no clique or $\ell$-clique component, there exists at
least one vertex $v$ adjacent to $M$. Since $M \cup \{v\}$ induces
a cluster subgraph in $G'$, all the $P_3$'s in $M \cup \{v\}$ are
destroyed by an optimal solution $F$. Therefore, if $|F|\leq k$ then $M$ contains at most $k+1$ vertices.

2) Exactly one $\ell$-clique $L$ and some cliques of size at least
two contain vertices of $M$. Let $C$ be one of these cliques. The
vertices of a P-vertex cannot be split into distinct parts of a
same $\ell$-clique. Therefore, exactly one vertex of $L$ contains
vertices of $M$ in $G'_{\mathrm Q}$. Let $C_M = C \cap M$ and $L_M
= L \cap M$. There exists at least one vertex $u \in C\backslash
C_M$ such that $u$ is adjacent to $M$ in $G$ (otherwise, we could
obtain a better solution if each vertex of $C_M$ was an isolated
clique in $G'$). Therefore, $u$ is adjacent to $L_M$ in $G$.
Similarly, there exists at least one vertex $v \in L\backslash
L_M$ such that $v$ is adjacent to $L_M$ in $G'$ and adjacent to
$M$ in $G$. Therefore, $v$ is adjacent to $C_M$ in $G$. Hence, the
edges $\{ul \mid l \in L_M\}$ and $\{vc \mid c \in C_M\}$
have been removed by an optimal solution $F$. If $|F|\leq k$ then
$|C_M|+|L_M|\leq k$. The same argument can be applied to other
cliques, if any. Therefore, $|M|\leq k$.

\begin{teo}
A problem kernel with $O(\ell k^2)$ vertices can be constructed
for $\L${\sc -Cluster Editing$(k)$} in $O(n+m)$ time.
\end{teo}

\emph{\textbf{Proof:}} By the previous analysis, we can construct
a problem kernel $G_k$ by restricting the size of the P-vertices
of $G_{\mathrm Q}$ to $k+2$ and the size of the S-vertices to
$\ell+k+1$. By Theorem \ref{sizeG_Q}, $G_k$ contains at most
$(2\ell k)(k+2)+(2k)(\ell +k +1) = O(\ell k^2)$ vertices. Graph $G_k$
can be constructed in $O(n+m)$ time by applying modular
decomposition \cite{DSPS2006,DSPS2009}. \hfill \framebox[.09in]

\section{Conclusions} \label{sec5}

The kernelization algorithms developed here and in \cite{DSPS2006,DSPS2009} can be applied to obtain, in linear time, special reduced graphs with $O(k)$ vertices which may help to solve {\sc Cluster Editing($k$)} and {\sc Bicluster Editing($k$)}, as explained below.

First, consider a generalization of {\sc Cluster Editing} (or {\sc Bicluster Editing}) in which edges and non-edges have positive integer weights (in the standard version, all edges/non-edges have weight one). The objective is then to obtain a cluster (bicluster) graph by applying to the input graph an edition set of minimum weight. The {\it weighted parameterized problem} associated with this generalization asks whether it is possible to obtain a cluster (bicluster) graph via an edition set of weight at most $k$. Let us denote it by {\sc Weighted Cluster Editing($k$)} ({\sc Weighted Bicluster Editing($k$)}).

Next, recall that if an instance $G$ of {\sc Cluster Editing($k$)} has answer `yes', then there exists an optimal solution such that no S-vertex $M$ of the S-quotient graph $G_{\mathrm S}$ is split into different vertices of $G'_{\mathrm S}$. Define weights for the edges of $G'_{\mathrm S}$ as follows: the weight of an edge $M\,M'$ of $G'_{\mathrm S}$ is the sum of the weights of all edges of $G$ with one endpoint in $M$ and other endpoint in $M'$ ($M$ and $M'$ can be modules of size larger than one). It is clear that $G$ is a yes-instance of {\sc Cluster Editing($k$)} if and only if $G'_{\mathrm S}$ (with the so-defined edge weights) is a yes instance of {\sc Weighted Cluster Editing($k$)}. Moreover, $G'_{\mathrm S}$ contains $O(k)$ vertices.

The same argument of the previous paragraph can be applied to {\sc Bicluster Editing($k$)} and the graphs $G_{\mathrm P}$ and $G'_{\mathrm P}$.

However, since in the problem $\L${\sc -Cluster Editing($k$)} P-vertices and S-vertices are in general unavoidably split into different vertices in an optimal solution, the Q-quotient graph $G_{\mathrm Q}$ cannot be used as above. In this case, the modular decomposition technique provides an $O(k^2)$ kernel in linear time.

A future work is the development of linear size kernels for $\L${\sc -Cluster Editing$(k)$}.

\bibliographystyle{plain}

\end{document}